\RequirePackage{cmap}
\documentclass[reprint,superscriptaddress,aps,prl,showemail,showpacs,footinbib]{revtex4-1}

\usepackage{amsthm}
\usepackage{amsmath}
\usepackage{latexsym}
\usepackage{amsfonts}
\usepackage{amssymb}
\usepackage{color}
\usepackage{bbm,dsfont}
\usepackage{graphicx}
\usepackage{enumerate}
\usepackage{hyperref}
\usepackage{subfigure}
\usepackage{tikz}
\usepackage{placeins}
\usepackage[utf8]{inputenc}
\usepackage{ulem}
\newcommand\opone{\leavevmode\hbox{\small1\kern-3.8pt\normalsize1}}
\newcommand{\e}{{\rm e}}
\DeclareUnicodeCharacter{00A0}{ }
 
\newtheorem{proposition?}{Proposition?}

\theoremstyle{definition}

\newcommand{\blue}[1]{\textcolor{blue}{#1}}

\newcommand{\real}{\mathbb R} 
\newcommand{\complex}{\mathbb C} 
\newcommand{\integer}{\mathbb Z} 

\newcommand{\hi}{\mathcal{H}} 
\newcommand{\ket}[1]{|#1\rangle} 
\newcommand{\kb}[2]{|#1\rangle\langle#2|} 
\newcommand{\tr}[2]{\text{tr}_{#2}\left\{#1\right\}}

\newcommand{\id}{\mathbbm{1}} 
\newcommand{\fii}{\varphi}

\newcommand{\va}{\mathbf{a}} 
\newcommand{\ve}{\mathbf{e}} 
\newcommand{\vu}{\mathbf{u}} 
\newcommand{\vk}{\mathbf{k}} 
\newcommand{\vx}{\mathbf{x}} 
\newcommand{\vsigma}{\boldsymbol{\sigma}} 

\newcommand{\mattt}[1]{\left( \begin{array}{ccc} #1 \end{array} \right)} 

\newcommand{\diff}{{\rm d}}
\newcommand{\norm}[1]{\vert\vert #1\vert\vert}
\newcommand{\abs}[1]{\vert #1 \vert}

\newcommand{\U}{{\mathcal U}}
\newcommand{\cyc}{\mathsf{cyc}}

\begin{document}
\title{Experimental and theoretical 
characterization  of a non-equilibrium steady state of a 
periodically driven qubit}

\author{Yong-Nan Sun}
\affiliation{CAS Key Laboratory of Quantum Information,
University of Science and Technology of China, Hefei, 230026, China}
\affiliation{CAS Center for Excellence in Quantum Information and Quantum Physics, University of Science and Technology of China, Hefei, 230026, P.R. China}

\author{Kimmo Luoma}
\email{ktluom@utu.fi}
\affiliation{Turku Center for Quantum Physics, Department of Physics and Astronomy,
University of Turku, FI-20014 Turun yliopisto, Finland}
\affiliation{Institut f{\"u}r Theoretische Physik, Technische Universit{\"a}t Dresden, 
D-01062,Dresden, Germany}

\author{Zhao-Di Liu}
\affiliation{CAS Key Laboratory of Quantum Information,
University of Science and Technology of China, Hefei, 230026, China}
\affiliation{CAS Center for Excellence in Quantum Information and Quantum Physics, University of Science and Technology of China, Hefei, 230026, P.R. China}

\author{Jyrki Piilo}
\email{jyrki.piilo@utu.fi}
\affiliation{Turku Center for Quantum Physics, Department of Physics and Astronomy,
University of Turku, FI-20014 Turun yliopisto, Finland}

\author{Chuan-Feng Li}
\email{cfli@ustc.edu.cn}
\affiliation{CAS Key Laboratory of Quantum Information,
University of Science and Technology of China, Hefei, 230026, China}
\affiliation{CAS Center for Excellence in Quantum Information and Quantum Physics, University of Science and Technology of China, Hefei, 230026, P.R. China}

\author{Guang-Can Guo}
\affiliation{CAS Key Laboratory of Quantum Information,
University of Science and Technology of China, Hefei, 230026, China}
\affiliation{CAS Center for Excellence in Quantum Information and Quantum Physics, University of Science and Technology of China, Hefei, 230026, P.R. China}




\date{\today}


\begin{abstract}
{Periodically driven dynamics of open quantum systems is very interesting because 
typically non-equilibrium
steady state is reached, which is characterized by a non-vanishing 
current.}
In this work, we study time discrete and periodically driven
dynamics experimentally for a single photon
that its coupled to its environment. {We develop a comprehensive theory which
explains the experimental observations and  offers an analytical characterization 
of  the 
non-equilibrium steady states of the system.} 
We {demonstrate} that the periodic driving and the
properties of the environment can be engineered in such a way that
there is asymptotically non-vanishing {bidirectional} information flow between
the open system and the environment.  

\end{abstract}

\maketitle

\paragraph{Introduction.---}
In realistic situations quantum systems are not  
isolated from the surrounding environment~\cite{alicki1987quantum}.
The coupling of the open system to the external 
world dynamically creates correlations between
the open system and it's environment. These then
lead to typically adversary effects such as 
decoherence and dissipation on the
level of the reduced state of the open system~\cite{breuer2002theory}. 

To cope with this unavoidable interaction one can try 
to perform quantum operations 
on a time scale that is faster than the decoherence and 
dissipation timescale~\cite{1367-2630-16-5-053049,Nielsen:2011:QCQ:1972505}, 
so that negative environmental effects do not degrade the quantum state.
Another route is to protect the quantum state is 
by using reservoir engineering to modify the 
environment \cite{2008NatPh...4..878D} or to do some local control operations
on the open system in hope to drive the system 
into decoherence free subspace \cite{PhysRevA.80.052316},
for example.
Recent years, significant technological advancements
have made it possible to realize
both strategies in various different experimental platforms, 
such as,  ion traps
~\cite{Schaefer_2018,1367-2630-16-5-053049,PhysRevLett.112.190502}, 
NV-centers~\cite{2014NatPh..10..720P} and 
single photons in free space~\cite{2017arXiv171208071L}, to
name a few.

From both experimental and theoretical point of view, 
{\it periodical}
local control strategies, leading to so called Floquet
dynamics \cite{PhysRevLett.124.170602,PRX9.021037}, for open quantum systems are very interesting. 
For example,
the open system typically will not reach a 
stationary state 
but rather the local driving forces the system asymptotically
to a non-equilibrium steady state, which is characterized
by non-vanishing current \cite{PhysRevLett.102.207207,2017NatPh..13..460B}.
Another interesting class of system, where effects similar to
a periodical driving occur are discrete 
time quantum walks \cite{PhysRevA.48.1687}, where the
time discrete nature of the dynamics leads to the 
Floquet theory and interesting phenomena ranging 
from topologically protected modes \cite{PhysRevA.82.033429} to
controlled transition from ballistic spread to 
localization \cite{PhysRevLett.106.180403} can be 
observed. Hardly any analytical results characterizing the 
asymptotic non-stationary steady states occurring in these
generally driven and dissipative systems exist.

In this Letter we study time discrete and periodically driven dynamics
both theoretically and experimentally for a single photon that its coupled 
to its environment. The polarization degrees of freedom of the photon
act as an open system and the frequency degree of freedom will serve 
as the environment, as in~\cite{Liu2011,2017arXiv171208071L}
We investigate the long time dynamics dynamics of the 
system experimentally and reveal signatures of the non-equilibrium
steady state from the data based on the comprehensive theory we 
develop. In particular, we show that the periodic driving and
the properties of the environment can be engineered in such a way
that there is asymptotically non-vanishing bidirectional exchange 
of information between the open system and the environment
leading to unbounded non-Markovianity. 

\paragraph{{Experimental setup---}}
\begin{figure}[t]
\includegraphics [width= 3.5 in]{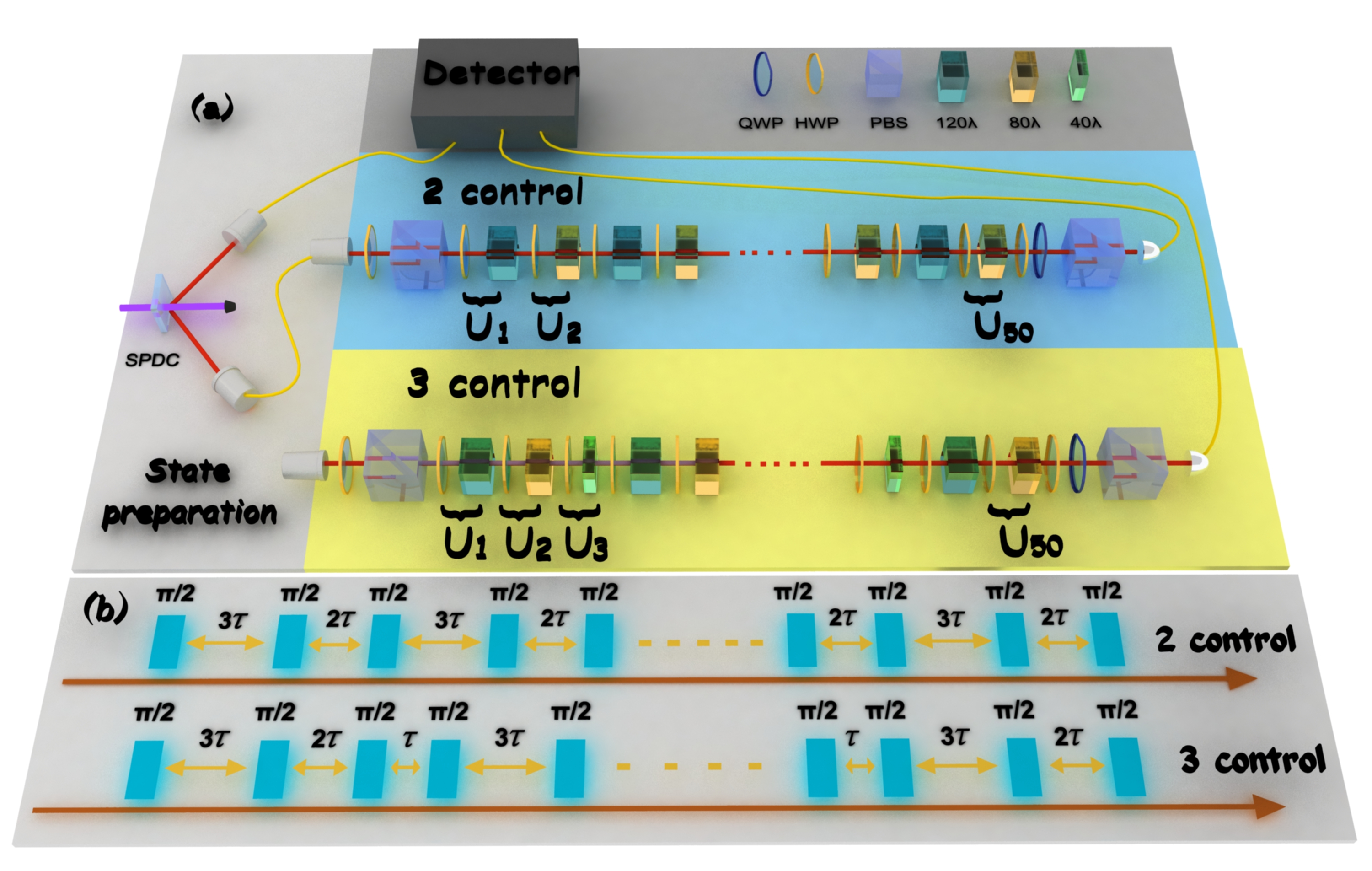}
\caption{{Experimental setup for testing the quantum 
    non-equilibrium steady state. (a) In our experiment, we realize the 
    unitary control and decoherence with a half-wave plate (HWP) and a 
    quartz plate which is called a operation unit $U$. There are in total 
    of 50 sets of operation units in our experiment. (b) Schematic picture 
    of the experiment. We show two dynamic progresses leading to the 
    non-equilibrium steady state which we call two controls 
    and three controls in our experiment. Two kinds of operation units 
    are used in the case of two controls to realize the non-equilibrium 
    steady state. For the case of three controls, three kinds of operation 
    units are used. Legend: QWP, quarter-wave plate; 
    HWP, half-wave plate; PBS, polarizing beam splitter; 
    3 kinds of quartz plates with different effective path difference: 
    $\Delta L=120\lambda$, $\Delta L=80\lambda$ and $\Delta L=40\lambda$.}} 
\label{fig:experimental_setup}
\end{figure}

{Our experimental setup is shown in Fig. 1.  In our experiment, a
type-\uppercase\expandafter{\romannumeral2} beta-barium-borate (BBO,
$9.0 \times 7.0 \times 1.0$ $mm^{3}$, $\theta = 41.44^{\circ}$)
crystal is pumped by a frequency-doubled femtosecond pulse (400 nm, 76
MHz repetition rate) from a mode-locked Ti:sapphire laser to generate
the degenerate photon pairs. After passing through the interference
filter (IF, $\Delta \lambda$ = 3 nm, $\lambda$ = 800 nm), the photon
pairs generated in the spontaneous parametric down conversion (SPDC)
process are coupled into single-mode fibers separately.  A
single-photon state is prepared by triggering on one of the two
photons, and the coincidence counting rate collected by the avalanche
photodiodes (APDs) is about $1.8 \times 10^{5}$ in 60 s.}

{As shown in Fig. 1, the single photon states are initialized into
state $\vert H \rangle$ by the first polarizing beam splitter. We use
a half-wave plate (HWP) and a quartz plate to realize the unitary
control and decoherence. The combination of a half-wave plate and a
quartz plate is called a operation unit $U$. There are in total of 50 sets
of operation units in each of our experiment. The refraction indices $n_{H}$ and
$n_{V}$ of the quartz plates are 1.5473 and 1.5384, respectively. We
have three kinds of quartz plates with different thicknesses,
corresponding to different dephasing strengths. We express the
thicknesses of the quartz plates in terms of effective path
difference, $\Delta L=40\lambda$, $\Delta L=80\lambda$ and $\Delta
L=120\lambda$, which correspond to the thicknesses of 3.556 mm, 7.111
mm and 10.667 mm, respectively. Different angles of the half-wave plates
lead to different values of the parameter $\eta$ and different local control
schemes. The value  $\eta=0.5$ use in each operation unit $U_i$ 
corresponds to a unitary rotation of the polarization of the photon effected 
by a half-wave plate set in angle of $22.5^{\circ}$ incident to the photon. 
At the final step, we make a
quantum state tomography for the single photon states after each step
with a polarizing beam splitter, a half-wave plate and a quarter-wave
plate.}

{For the case of two controls, two kinds of quartz plates are used,
$\Delta L=120\lambda$ and $\Delta L=80\lambda$ for the 
operation units $U_1$ and $U_2$, respectively.  
We stack in total 50 operation units cyclically $U_2U_1\cdots U_2U_1$ 
in our experiment.}

{For the case of three controls, three kinds of quartz plates are used.
In operation unit $U_1,\, U_2$ and $U_3$  the thicknesses are  
$\Delta L=120\lambda$, in $U_2$ $\Delta L=80\lambda$ and $\Delta
L=40\lambda$, respectively. 
We stack in total 50 operation units cyclically $U_2U_1U_3U_2U_1\cdots U_3U_2U_1$
in our experiment. Note here that $50\mod 3 = 2$ and therefore the $50$th 
operation unit is $U_2$.}

\begin{figure}[h!]
  \includegraphics[keepaspectratio,width=0.46\textwidth]{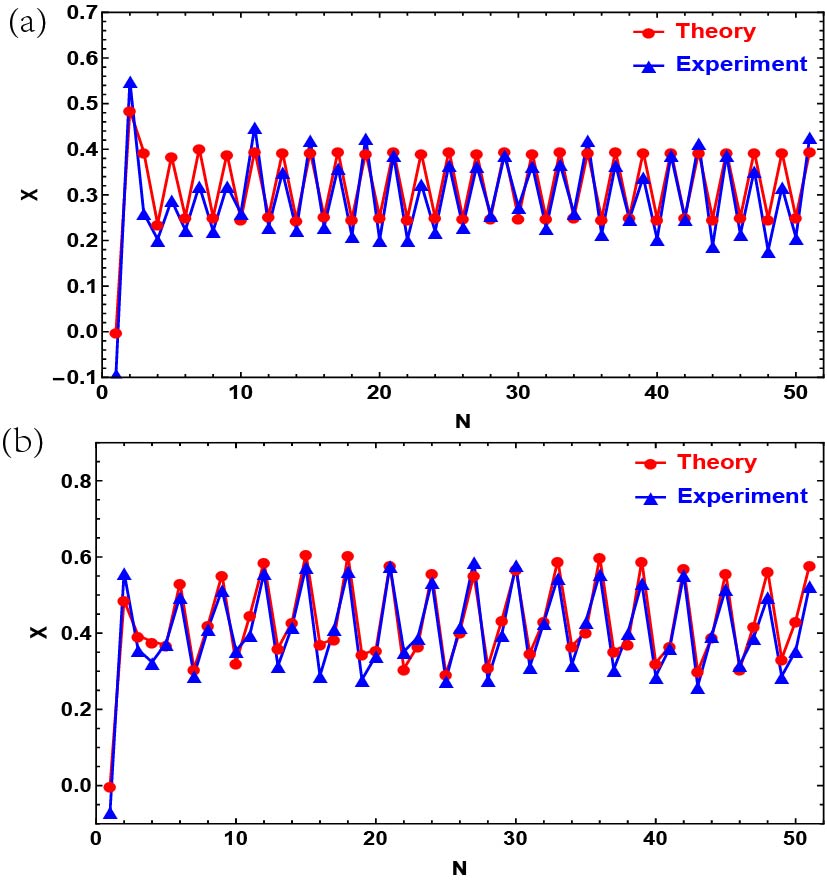}
  \caption{\label{fig:experimental_results1} Evolution of the
polarization state of the photon for the initial state
$\ket{\varphi}=\ket{H}$ after 50 steps. (a)Theoretical and
experimental results for two controls. (b) Theoretical and
experimental results for three controls.}
\end{figure}

\begin{figure}[h!]
  \includegraphics[keepaspectratio,width=0.44\textwidth]{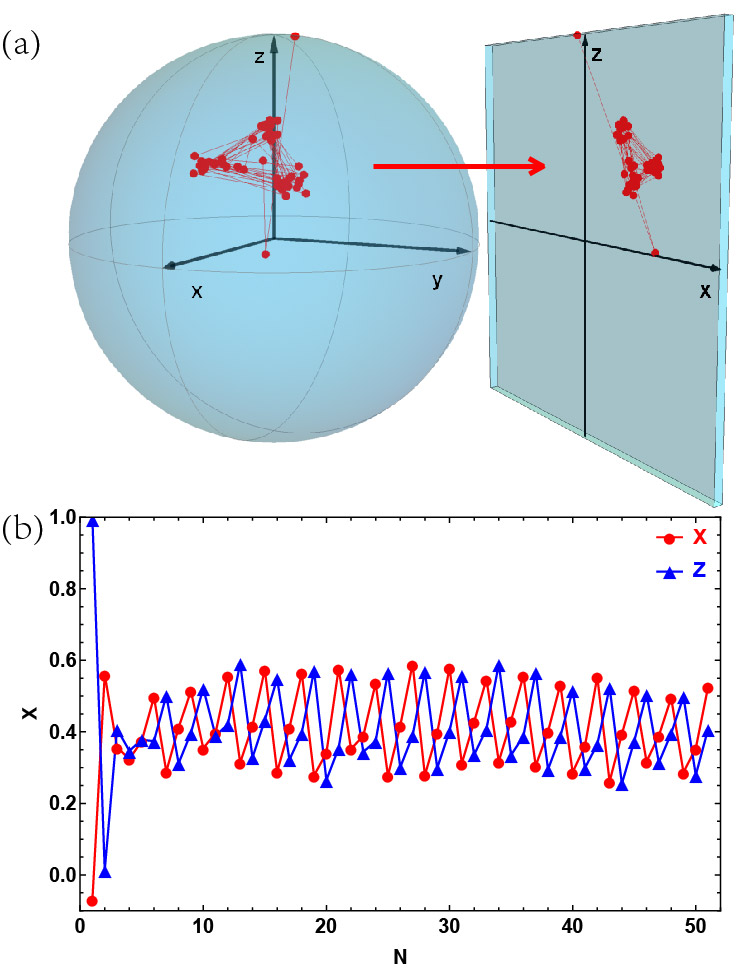}
  \caption{\label{fig:experimental_results2} Evolution of the 
    polarization state of the photon for the initial state
$\ket{\varphi}=\ket{H}$ for three controls. (a) Evolution of the
polarization state in bloch sphere. (b) Dynamics for the x and z
components of the bloch vector for three controls.}
\end{figure}

\paragraph{Experimental results.---}
{In the case of two controls, we have initialized the polarization to state 
$\ket{\fii}=\ket{H}$. After short initial transient the polarization oscillates 
between two values, as can be seen in Fig.~\ref{fig:experimental_results1}
a). Our theoretical prediction fits well to the experiment. In panel b) of 
Fig.~\ref{fig:experimental_results1} we use three controls and find 
a good match between the theory and experiment. In the case of three
controls the asymptotic dynamics oscillates between three different 
polarization states as can be seen from Fig.~\ref{fig:experimental_results2}
a) and b). In both the two and three control cases the dynamics is confined to the plane
$y=0$ cutting the Bloch sphere by a suitable choice of the initial state. Interesting
features of this rather simple looking dynamics can be revealed from the 
theory developed next.}

\paragraph{Theory.---}
Quantum mechanical state of a single photon wave packet 
is an element of $\hi=\hi_S\otimes\hi_E=\complex^2\otimes L^2(\real)$,
where the polarization and the frequency of the photon 
correspond to the first and to the second Hilbert space, respectively.
The photon is prepared initially into a pure state 
$\ket{\Psi}=\int\diff\omega\, \chi(\omega)\ket{\varphi}\otimes\ket{\omega}$,
in such a way, that 
the {frequency} spectrum of the photon  is \blue{a} Gaussian
\begin{align}\label{eq:spectra}
  \abs{\chi(\omega)}^2
        =\frac{1}{\sqrt{2\pi}\sigma}e^{-(\omega-\mu)^2/2\sigma^2},
\end{align}
with central frequency $\mu$ and standard deviation $\sigma$.
The photon 
propagates trough a set of linear optical elements
that sequentially rotate its polarization state and couple 
the polarization and frequency degrees of freedom. 
Such a procedure defines the following dynamical map 
that acts on the polarization state of the photon
\begin{subequations}
  \begin{align}\label{eq:dyn_map}
    \Phi_n[\rho_0]=&\int_\real\diff\omega\, \abs{\chi(\omega)}^2
                     \prod_{i=0}^{n-1}\U_{n-i}(\omega)[\rho_0],\\
    \U_i(\omega)[\rho]=& 
                         U_i(\omega)\cdot(C_{\eta_i}\otimes\id_E)\rho
                         (C_{\eta_i}\otimes\id_E)^\dagger\cdot U_i(\omega)^\dagger
  \end{align}
\end{subequations}
where $U_i(\omega)= e^{i n_1\omega L_i/c}\kb{1}{1}+e^{i n_2\omega L_i/c}\kb{2}{2}$ couples
the polarization and frequency degrees of freedom and 
$C_{\eta_i}=\sqrt{\eta_i}\sigma_z +\sqrt{1-\eta_i}\sigma_x$ rotates 
the polarization. $0\leq \eta_i\leq 1$ parametrize the rotations, 
$n_i$,$L_i$ and $c$ parametrize the couplings between 
the open system and it's environment corresponding to
the indices of refraction and length of a quartz plate where $c$ is 
the speed of light. We immediately see that  
the map is unital $\Phi_n(\id)=\id$. From now on we set 
$\eta_i=\eta$ for each step and $\eta = 0.5$ whenever numerical values are required.

In the Bloch picture $\rho=\frac{1}{2}(\id+\va\cdot \vsigma)$ 
the unitary operators are 
\begin{subequations}
  \begin{align}
    \U(\omega)=&\mattt{\beta \cos\frac{\Delta n\omega L}{c} 
    &-\beta\sin\frac{\Delta n \omega L}{c}
    &\alpha\\
    -\sin\frac{\Delta n \omega L}{c} 
               & -\cos\frac{\Delta n \omega L}{c} 
    & 0\\
    \alpha\cos\frac{\Delta n \omega L}{c} 
               &-\alpha\sin\frac{\Delta n \omega L}{c} 
    &-\beta}\\\label{eq:5}
    C_\eta=&\mattt{\beta & 0 & \alpha\\ 0 & -1 & 0\\
    \alpha & 0 & -\beta},\\\label{eq:6}
    U(\omega)=&\mattt{\cos \frac{\Delta n \omega L}{c} 
    &-\sin\frac{\Delta n \omega L}{c} & 0 \\
    \sin \frac{\Delta n  \omega L}{c} 
               & \cos \frac{\Delta n \omega L}{c} & 0 \\
    0 & & 1}
  \end{align}
\end{subequations}
where $\Delta n = n_1-n_2$, $\beta = 1-2\eta$ and $\alpha = 2\sqrt{(1-\eta)\eta}$. 
We see that the 
unitary operators are mapped to orthogonal matrices acting on
$\real^3$. Initial Bloch vector $\va$ is mapped to 
$\va_n=\int\diff\omega\, \abs{\chi(\omega)}^2\prod_{i=0}^{n-1}\U_{n-i}(\omega)\va.$
Further, we make a change of variables 
$t=\tan\frac{\Delta n L\omega}{c}$, such that the trigonometric
functions are mapped to polynomials in $t$. The details of this 
transformation are in 
the Supplementary Material~\cite{Supplement}. 

Open system dynamics leading to non-equilibrium steady state
can be engineered if we periodically vary the linear optical
elements. Let $T\in\integer_+$ be a length of such a period.
The asymptotic
dynamical maps $\tilde\Phi_{Tm+K}$, where $0\leq K < T$ can be seen as a
integer valued phase of the periodic driving, characterize the asymptotic
dynamics. 
Maps $\tilde\Phi_{Tm+k}$ are obtained using generating function 
techniques and asymptotic analysis \cite{Flajolet:2009:AC:1506267}, which
we now present.


Let $\{H_n\}_{n=0}^\infty$ be a family of 
orthogonal matrices acting on $\real^3$. 
Further, matrices $H_n$ satisfy the following recursion 
relation
\begin{align}
  \label{eq:recursion}
  H_{n+1}=W H_n,\,\,\forall n\geq 0, &&H_0 = A,
\end{align}
where $A$ is arbitrary initial condition and 
$W$ is an orthogonal matrix. Then we 
define the $z$-transform as $H(z)=\sum_{n=0}^\infty z^nH_n$, where 
$z\in\complex$.

Multiplying both sides of Eq.~(\ref{eq:recursion}) with 
$z^n$ and performing the sum $\sum_{n=1}^\infty$ formally on both 
sides, we obtain 
\begin{align}\label{eq:z-transform}
  H(z)=&(\id-zW)^{-1}A.
\end{align}
By Cramers rule we know that 
\begin{align}\label{eq:Cramers}
  (\id-z W)^{-1}=\frac{\mathsf{adj}(\id-z W)}{\mathsf{det}(\id-z W)},
\end{align}
where $\mathsf{adj}(\cdot)$ denotes the adjungate matrix. 
$W$ is orthogonal matrix acting on $\real^3$ and therefore
it has eigenvalue $\lambda=1$. Since determinant 
is a product of eigenvalues, we see that 
all the matrix elements of $H(z)$ have a simple 
pole at $z=1$. 
On the other hand, we have the following 
asymptotic relation \cite{Flajolet:2009:AC:1506267}
\begin{align}\label{eq:zto1}
  \lim_{n\to\infty}W^nA\sim\lim_{z\to 1}(1-z)H(z)=\mathsf{Res}(H(z),z=1),
\end{align} 
which can be easily seen by expanding the geometric
series and telescoping the resulting expression.
The $\mathsf{Res}(H(z),z=z_0)$ corresponds to 
element-wise residue operation.

Suppose now that we have a situation where the matrix 
$H_n$ satisfies a periodic recursion relation with 
period $T$. First we define an operator
\begin{align}
 \cyc(W_TW_{T-1}\cdots W_1)=W_{1}W_{T}\cdots W_2,
\end{align}
that is $\cyc$ shifts the operators cyclically to 
the right. For example
$\cyc^2(ABC)=\cyc(CAB)=BAC$ and so on. 
Then we can write for any $0 \leq K < T$ 
\begin{align}\label{eq:periodic_evol}
  H_{mT+K}=&\cyc^K(W_T\cdots W_1)^mW_KW_{K-1}\cdots W_1A.
\end{align}
Meaning of the above formula is easy to understand. 
Integer $K$ corresponds to the phase of the evolution 
over a single period of length $T$. 

This observation opens up a possibility to have 
maps that do not have a stationary state but rather a
non-equilibrium steady state. This can 
be seen by applying Eq. (\ref{eq:z-transform}) to the 
periodically driven evolution (\ref{eq:periodic_evol}). Namely, if  
there exist a set of integer phases $S$, such that 
the asymptotic limit maps differ, ie.
\begin{align}
  &\lim_{n\to\infty}\cyc^{K}(W_T\cdots W_1)^nW_{K}\cdots W_1A \notag \\
  &\neq  \lim_{n\to\infty}\cyc^{K'}(W_T\cdots W_1)^nW_{K'}\cdots W_1A,
\end{align}
for any $K',K'\in S$, then asymptotically we have 
$\abs{S}$ phase dependent limit maps. $\abs{S}$ corresponds
to the number of elements in the set $S$.


\paragraph{{Application of the general theory to two and three controls---}}
For the case of two controls, {for which 
$(\Delta n L_1)/(\Delta n L_2)=3/2$, where $\Delta nL_i$ is the dephasing strength and $\eta_{1}=\eta_{2}=1/2$}, we find that the 
asymptotic dynamical map takes the following
form
\begin{subequations}
  \begin{align}
    \label{eq:two_cont_0}
    \tilde\Phi_{2m}
    &=\mattt{
      0.635946 & 0 & 0.394485 \\
    0 & 0.114589 & 0 \\
    0.394485 & 0 & 0.249465},\\
    \label{eq:two_cont_1}
    \tilde\Phi_{2m+1}
    &=\mattt{
     0.394485 & 0 & 0.249465 \\
    0 & 0.114589 & 0 \\
    0.635946 & 0 & 0.394485}.
  \end{align}
\end{subequations}
In this case, all eigenvalues of 
both asymptotic maps are non-zero and non-degenerate. 
Initial states with Bloch vector $\va=a\ve_y$, with  
$\abs{a}\leq 1$ map asymptotically to 
$\va\mapsto \lambda_y^{(2m)}a\ve_y=\lambda_y^{(2m+1)}a\ve_y$, 
which is the same state
for even and odd number of steps because
$\lambda_y^{(2m)}=\lambda_y^{(2m+1)}=0.114589$. 
The two dimensional subspace 
orthogonal to $\ve_y$ is spanned by vectors 
$\vx_1^{(2m)}$ and $\vx_3^{(2m)}$, which correspond to 
first and third columns
of the matrix $\tilde\Phi_{2m}$. 
In this subspace the asymptotic dynamics corresponds to $\sigma_x$ i.e.
$\vx_{1/3}^{(2m+1)}=\sigma_x\vx_{1/3}^{(2m)}$. We can conclude 
that in the two control case (i) the purity stays asymptotically
constant since $\sigma_x$ is unitary and (ii) 
the initial states that have support in the 
subspace orthogonal to $\ve_y$ oscillate asymptotically
with the same period as the driving. 
{The asymptotic ``flipping'' dynamics is clearly visible in Fig.~\ref{fig:experimental_results1} a).}

For the case of three controls, which {satisfy 
$(\Delta nL_1)/(\Delta n L_2)=3/2$, $(\Delta n L_2)/(\Delta n L_3) =2$ and $\eta_{1}=\eta_{2}=\eta_{3}=1/2$}, we find that the 
asymptotic propagators have the following structure
\begin{subequations}
  \begin{align}
    \label{eq:three_cont_3}
    \tilde\Phi_{3m}(t)
    =&\mattt{
       0.363253 & 0 & 0.331023 \\
    0 & 0.0590277 & 0 \\
    0.331023 & 0 & 0.577719},\\
    \label{eq:three_cont_1}
    \tilde\Phi_{3m+1}(t)
    =&\mattt{
       0.350767 & 0 & 0.399416 \\
    0 & 0.127151 & 0 \\
    0.363253 & 0 & 0.331023},\\
    \label{eq:three_cont_2}
    \tilde\Phi_{3m+2}(t)
    =&\mattt{
       0.331023 & 0 & 0.577719 \\
    0 & -0.0386657 & 0 \\
    0.350767 & 0 & 0.399416}.
  \end{align}
\end{subequations}
The asymptotic maps 
$\tilde\Phi_{3m+i}$, $0\leq i < 3$ can be all decomposed
to one dimensional subspace spanned by $\ve_y$ and 
its orthogonal complement. We denote the eigenvalues 
as $\lambda_y=\lambda_y^{(3m)}$, $\lambda_y'=\lambda_y^{(3m+1)}$ and 
$\lambda_y''=\lambda_y^{(3m+2)}$.   We have $\abs{\lambda_y''}<\abs{\lambda_y}<\abs{\lambda_y'}$, which
shows that there exists initial states whose purity
changes asymptotically. Namely, initial states given by 
Bloch vector $\va=a\ve_y$, where $\abs{a}\leq 1$.
The orthogonal complements are spanned by vectors 
$\vx_{1/3}$, $\vx_{1/3}'$ and $\vx_{1/3}''$ corresponding 
to the first- and third column of maps 
~(\ref{eq:three_cont_3}),(\ref{eq:three_cont_1}) and 
(\ref{eq:three_cont_2}). Suppose we consider initial state
$\va=(1/\sqrt{2})(\ve_x+\ve_z)\perp\ve_y$. We see that 
$\norm{\tilde\Phi_{3m+1}\va}<\norm{\tilde\Phi_{3m}\va}<\norm{\tilde\Phi_{3m+2}\va}$,
thus we see that the purity changes also for vectors 
in the orthogonal complement of $\ve_y$. 
{The period of three steps is clearly visible in Fig.~\ref{fig:experimental_results1}
b) and the evolution of purity in Fig.~\ref{fig:experimental_results2} b).}

\paragraph{Maximal visibility.---}
To experimentally observe the non-stationary 
steady state, the initial state should be chosen 
such that its visibility is maximal. This means that
the volume spanned by the points in the limit cycle 
is maximal. 
{The volume is expressed as a}  
function $f_V(\va)=f_V(\theta,\varphi)$, where 
$\va=(\cos\varphi\sin\theta,\sin\varphi\sin\theta,\cos\theta)^T$ is the 
Bloch vector of the initial state. The extremal solutions
are found when $\partial_\theta f_V=\partial_\varphi f_V=0$. The
extremals correspond to maximums if the Hessian 
$H(f_V)$, that is the matrix formed from the 
second order partial derivatives, is negative definite.

For two point cycle, the volume is the Euclidean 
distance between the two asymptotic states
  $f_V^{(2)}(\theta,\varphi)= \norm{(\Phi_{2m}-\Phi_{2m-1})\va}^2$,
which is to be maximized.

For three point cycle the {volume is the area} spanned by the parallelogram formed by
{the} three asymptotic endpoint vectors. To compute it,
we first label the 
asymptotic states as 
$\vx=\Phi_{3m}\va$, $\vx'=\Phi_{3m+1}\va$ and 
$\vx''=\Phi_{3m+2}\va$. Then we define $\vu=\vx'-\vx$ and 
$\vk=\vx''-\vx'$. Now the {area} spanned by the parallelogram
is $\norm{\vu\times\vk}$ which reads  
  $f_V^{(3)}(\theta,\varphi)=\frac{1}{2}\norm{\vx\times\vx'+\vx'\times\vx''+\vx''\times\vx}$. 
When given the asymptotic maps the volumes 
$f_V^{(2)}$ and $f_V^{(3)}$ are very easy to 
maximize. {We present theoretical and experimental results for the maximal visibility
initial state in the Supplemental Material~\cite{Supplement}.}

\paragraph{Maximal non-Markovianity.---}
The non-stationary steady state can lead to 
unbounded non-Markovianity of the dynamics measured by 
the BLP-measure \cite{PhysRevLett.103.210401}. 
For example, in the case of asymptotic three cycle,
Eqs.~(\ref{eq:three_cont_3}-\ref{eq:three_cont_2}),
we consider pure state pairs initially polarized
in the $y$-direction
$\rho_y^\pm=\frac{1}{2}(\id\pm\ve_y\vsigma)$, for which $D(\rho_y^+,\rho_y^-)=1$. 
We use the following shorthand notation 
$D_m^y=D(\rho_y^+(m),\rho_y^-(m))$.
We see that 
asymptotically  
\begin{align}
D_{3m+1}^y-D_{3m}^y&=\abs{\lambda_y'}-\abs{\lambda_y} > 0,\notag\\ 
\label{eq:nm_example}
D_{3m+2}^y-D_{3m+1}^y&=\abs{\lambda_y''}-\abs{\lambda_y'} < 0,\\
D_{3m}^y-D_{3m+2}^y&=\abs{\lambda_y}-\abs{\lambda_y''} > 0.\notag 
\end{align}
The asymptotic growth emerges from the first and the 
second term.
The trace 
distance can be written as a maximization over 
all positive operators 
\begin{align}\label{eq:trace_dist}
  D(\rho_1,\rho_2)=\max_{F'}\tr{F'(\rho_1-\rho_2)}{},
\end{align}
$0< F\leq \id$ \cite{Nielsen:2011:QCQ:1972505}. In the 
running example the positive operator 
that maximizes Eq.~(\ref{eq:trace_dist}) for all
points in the asymptotic cycle corresponds to either
of the POVM's
$F_*'=\frac{1}{2}(\id\pm\e_y\vsigma)$. Therefore, the
lower bound for non-Markovianity is obtained from the 
asymptotic oscillations of the polarization in the 
$y$-direction.

Largest asymptotic growth of non-Markovianity is obtained for
orthogonal initial state pairs that asymptotically have largest change in 
the purity during one cycle. 
The proof of the claim is the following.
For qubit states the trace distance is proportional to 
the Euclidean norm of Bloch vector corresponding to 
the difference state $D(\rho_1,\rho_2)=\frac{1}{2}\norm{\va_1-\va_2}$.
The non-Markovianity measure is maximized for orthogonal
initial state pair, for which the initial Bloch vectors
are anti-podal, $\va_0^{\pm}$ \cite{PhysRevA.86.062108}. Since, the dynamical map is 
linear, the evolution of the anti-podal state pair 
has a reflection symmetry, i.e. $\va_m^+=-\va_m^-$.
Therefore, $D_m^\pm=D(\va_m^+,\va_m^-)=\norm{\va_m^\pm}$.
Length of the Bloch vector corresponds to the purity of 
the state. Then clearly, we have largest asymptotic
growth of non-Markovianity for initial states that 
have largest purity changes asymptotically. In general
the different asymptotic states during one cycle 
are not parallel i.e. $\va_m\nparallel\va_{m+k}$.
Therefore we can not find a single observable 
that would give the trace distance for the optimal
initial state at any point in the asymptotic cycle.

\paragraph{Conclusions.---}
Hardly any analytical results characterizing the asymptotic
non-stationary steady states occurring in these generally driven and
dissipative systems exist. We have studied time discrete and
periodically driven dynamics both theoretically and experimentally for
a single photon that its coupled to its environment. We have characterized
analytically the non-equilibrium steady states and experimentally
verified our theoretical results. We show that the periodic driving and
the properties of the environment can be engineered in such a way that
there is asymptotically non-vanishing exchange of information between
the open system and the environment, {i.e.  a current,
which is a hallmark of non-equilibrium steady-state and in this case
leads to asymptotically unbounded non-Markovianity.}

\paragraph{Acknowledgments.---}
This work was supported by the National Key Research and Development Program of
China (No. 2017YFA0304100), the National Natural Science Foundation of China (Nos.
61327901, 11325419, 11774335, 11821404), Key Research Program of Frontier Sciences,
CAS (No. QYZDY-SSW-SLH003), Anhui Initiative in Quantum Information Technologies (AHY020100), 
the Fundamental Research Funds for the Central Universities (No.
WK2470000026), Science Foundation of the CAS (No. ZDRW-XH-2019-1).

\bibliography{PeriodicallyDrivenSpinBib}
\end{document}